# Distributed Dynamic State Estimation for Microgrids

Bang L. H. Nguyen
ECE Department
Clarkson University
Potsdam, NY, USA
bangnguyen@ieee.org

Tuyen V. Vu
ECE Department
Clarkson University
Potsdam, NY, USA
tvu@clarkson.edu

Thomas H. Ortmeyer
ECE Department
Clarkson University
Potsdam, NY, USA
tortmeye@clarkson.edu

Tuan Ngo
Electric Power Engineers
Austin, TX, USA
ngotuan@utexas.edu

*Abstract*—**Conventionally, the dynamic state estimation of variables in power networks is performed based on the forecasting-aided model of bus voltages. This approach is effective in the stiff grids at the transmission level, where the bus voltages are less sensitive to variations of the load. However, in microgrids, bus voltages can fluctuate significantly under load changes, the forecasting-aided model may not sufficiently accurate. To resolve this problem, this paper proposes a dynamic state estimation scheme for microgrids using the state-space model derived from differential equations of power networks. In the proposed scheme, the branch currents are the state variables, whereas the bus voltages become the inputs which can vary freely with loads. As a result, the entire microgrids system can be partitioned into local areas, where neighbor areas share the common inputs. The proposed estimation scheme then can be implemented in a distributed manner. A novel Kalman-based filtering method is derived to estimate both states and inputs simultaneously. Only information of common inputs is exchanged between neighboring estimators. Simulation results of the 13-bus Potsdam microgrid (New York State) are provided to prove the feasibility and performances of the proposed scheme.**

*Index Terms*—**Bad data detection, dynamic state estimation, distributed scheme, data fusion, microgrids.**

## I. INTRODUCTION

Dynamic state estimation (DSE) is a crucial tool to perceive the real-time states of power systems by processing measurements and other information. Based on the accurate state estimates, optimal control, management, contingency analysis security assessment, and other applications can be performed appropriately [1]. Evolved from traditional static state estimation (SSE) [2], DSE with a filtering framework such as Kalman filter (KF) can keep track of the change of states and also utilize the historical data for estimation [3]. DSE can be implemented in either centralized or distributed fashions.

Since, in a centralized scheme, both computation and communication costs are high, distributed ones are preferred [4]. Many research efforts have been put on distributed filtering algorithms for applications from multi-sensors networks to large-scale systems [5]. In a distributed estimation scheme, local estimators can be executed independently in the same way as the centralized ones are with their internal measurements. Then, communication channels exchange local estimates information and the data fusion is performed to reconstruct the global state estimates.

Although distributed estimation can reduce both computational and communication burden, it could suffer from data consensus problems in which local estimates are diverged [6]. Hence, choosing an appropriate fusion architecture, which specifies communication and data fusion strategies, plays a key role in designing such that distributed state estimation system.

### A. Related Works.

The concept of parallel distributed estimation is developed from hierarchical state estimation (SE), which may be installed in a control center or distributed agents [7]. The standard distributed estimation procedure consists of two stages: 1) conventional SE is carried out independently in subareas with their local measurements, and 2) the obtained estimates are either exchanged between local estimators or collected to a center for the fusion of the final estimates. This procedure is still applicable to the static SE framework, where communication channels are assumed flawless.

The decentralized implementations for DSE are introduced in [8]; however, they ignore the global data fusion step. In [9], the authors implemented information filters in a decentralized manner, where information vectors and matrices of measurements are exchanged between local estimators. On the other hand, an equality-constrained optimization problem is employed in [10] to update the dynamic states with preprocessed corrective generator internal voltages and roto angles. However, these techniques are effective only under the assumption of good data and communication. [11] uses a consensus-based observer to obtain global estimates, which is resilient with missing data. However, the consensus algorithm requires certain time steps to converge. In addition, it can result in false alarm or bypassed bad-data before settling down. [12] uses the internodal transformation theory to sort out the related measurements from neighbors. Similarly, in [13] an inter-processor transformation matrix is employed to exchange and integrate the overlapped state estimates.

Nonetheless, using transformation matrices can result in a more complex computation, which should be lessened. In addition, all aforementioned studies, except [11], employ the forecasting-aided model of bus voltages to perform DSE. This approach is effective in the stiff grids at the transmission level, where the bus voltages are less sensitive to variations of the load. However, in microgrids, bus voltages can fluctuate significantly under load changes, the forecasting-aided model may not sufficiently accurate. Although [11] involves the dynamic model of microgrids, the authors considered all inputs in the state-space models are known exactly. This is not true due to the fact that all variables are measured with some noises.

*B. Contributions*

To resolve this problem, this paper proposes a dynamic state and input estimation scheme on the state-space models of microgrids. In the proposed scheme, the branch currents are the state variables, whereas the bus voltages become the inputs which can vary freely with loads. As a result, the entire microgrids system can be partitioned into local areas, where neighbor areas share the common inputs. The dynamic estimation then can be implemented in a distributed manner.

The main contributions of this paper are as follows: 1) This work pioneer to resolve the problem of unknown inputs in dynamic state estimation using state-space models, 2) A novel Kalman-based filtering method is derived to simultaneously estimates both states and inputs, 3) The distributed implementation of proposed estimation scheme is introduced. 4) The communication and bad data detection protocol are presented.

For the rest of the paper, Section II introduces a dynamic model of microgrids network and a centralized formulation of dynamic state and input estimation (DSIE). Section III implements the distributed DSIE with communication and bad data detection protocol. The simulation results are provided and discussed in Section IV. Section V concludes the paper.

## II. MICROGRIDS MODELS AND CENTRALIZED DSIE

*A. Microgrid Models*

A microgrid system includes distributed generation units (DGUs), power networks, and loads. As expressed in [14], each DGU with a connected inverter can be presented as a controllable voltage source as shown in Fig. 1. DGU's model can be presented in *dq*-frame as

$$\frac{di_{ti,dq}}{dt} + j\omega_o i_{ti,dq} = -\frac{R_{ti}}{L_{ti}} i_{ti,dq} + \frac{1}{L_{ti}}(v_{ti,dq} - k_i v_{i,dq}), \quad (1)$$

$$\frac{dv_{i,dq}}{dt} + j\omega_o v_{i,dq} = \frac{1}{C_{ti}}(k_i i_{ti,dq} + i_{ij,dq} - i_{Li,dq}), \quad (2)$$

where $i, j$ denote the bus index; $\bullet_{dq}$ denotes as $(\bullet_d + j\bullet_q)$; $\omega_o$ is the angular frequency of the *dq*-rotating frame.

The dynamic model of line *ij* is presented as

$$\frac{di_{ij,dq}}{dt} + j\omega_o i_{ij,dq} = -\frac{R_{ij}}{L_{ij}} i_{ij,dq} + \frac{1}{L_{ij}}(v_{j,dq} - v_{i,dq}). \quad (3)$$

For the bus with a capacitor bank or having sufficient shunt susceptances, its dynamic model can be described as

$$\frac{dv_{j,dq}}{dt} + j\omega_o v_{j,dq} = \frac{1}{C_j} \sum i_{j*,dq}, \quad (4)$$

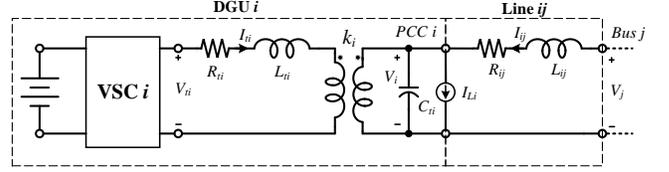

Figure 1. Single-line equivalent circuit of a DGU and a power

where $\sum i_{j*,dq}$ indicates the algebraic sum of all currents injected into bus *j*, except the capacitor current. If there is no capacitor at bus *j*, this sum should equal to zero.

From (1)-(4), the state equations of microgrids therefore can be expressed as

$$\dot{x} = Ax(t) + Bu(t) + w(t), \quad (5)$$

where the state vector $x \in \mathbb{R}^n$ contains all branch currents and voltages of buses with a capacitor, the input vector $u \in \mathbb{R}^m$ includes voltages of buses without capacitor or load currents, $A$ is the state matrix, $B$ is the input matrix, and $w(t)$ represents the process noise.

*B. Measurement Models*

The measurements can be obtained by micro-phasor measurement units (µPMU) which can directly provide synchronized currents and voltages phasors in rectangular form (*dq*-frame) as

$$z_x(t) = Cx(t) + v_x(t), \quad (6)$$

$$z_u(t) = Du(t) + v_u(t), \quad (7)$$

where $C$ and $D$ are state and input measurement matrices, respectively, $v_x$ and $v_u$ are the measurement noises, $z_x \in \mathbb{R}^p$, and $z_x \in \mathbb{R}^l$.

The power measurements are given as

$$P(t) = v_{*,d}(t)i_{*,d}(t) + v_{*,q}(t)i_{*,q}(t), \quad (8)$$

$$Q(t) = -v_{*,d}(t)i_{*,q}(t) + v_{*,q}(t)i_{*,d}(t), \quad (9)$$

where the subscript (∗)

The synchronized measurements introduce the linear output equation, while the power measurement equations are nonlinear. In this paper, we only consider measurements from µPMU, power measurements are assumed that can be converted to linearized ones as expressed in [16].

*C. Centralized Dynamic State and Input Estimation (DSIE)*

Discretizing (5) with the sampling time $T_s$ and combining with (6) and (7), the discrete state-space model of microgrids can be expressed as

$$\begin{cases} x_k = A_d x_{k-1} + B_d u_{k-1} + w_{k-1}, \\ z_{x,k} = Cx_k + v_{x,k}, \; z_{u,k} = Du_k + v_{u,k}. \end{cases} \quad (10)$$

where $k$ indicates the time step, $A_d$ and $B_d$ are the discrete state and input matrices, respectively [15]. The covariance matrices of process and measurement noises ($w, v_x, v_u$) are assumed as white and Gaussian noises having zero means and covariance matrices of ($Q, R_x, R_u$) respectively.

*Input Estimation*:

The input-to-output relationship of the system can be derived from (10) as

$$z_{x,k} = CA_d x_{k-1} + CB_d u_{k-1} + Cw_{k-1} + v_{x,k}. \quad (11)$$

Combining (11) with the previous state estimates and input measurements, the following set of equations can be achieved.

$$\begin{bmatrix} \hat{x}_{k-1} \\ z_{u,k-1} \\ z_{x,k} \end{bmatrix} = \begin{bmatrix} I & 0 \\ 0 & D \\ CA & CB \end{bmatrix} \begin{bmatrix} x_{k-1} \\ u_{k-1} \end{bmatrix} + \begin{bmatrix} e_{x,k-1} \\ v_{u,k-1} \\ Cw_{k-1} + v_{x,k} \end{bmatrix}, \quad (12)$$

where $\hat{x}_{k-1}$ is the state estimates of time step $(k-1)$ and $e_{x,k-1}$ is the estimated error with covariance matrix $P_{x,k-1}$. Since $I$ is the identity matrix, (12) is overdetermined when $rank\left(\begin{bmatrix} I & 0 \\ 0 & D \\ CA & CB \end{bmatrix}\right) \geq m$. Given $\mathcal{O} = \begin{bmatrix} I & 0 \\ 0 & D \\ CA & CB \end{bmatrix}$, the WLS solution of (12) can be expressed as

$$\begin{bmatrix} \hat{x}_{k-1} \\ \hat{u}_{k-1} \end{bmatrix} = (\mathcal{O}^T R^{-1} \mathcal{O})^{-1} \mathcal{O}^T R^{-1} \begin{bmatrix} \hat{x}_{k-1} \\ z_{u,k-1} \\ z_{x,k} \end{bmatrix}, \quad (13)$$

where the weighted matrix $R = diag(P_{x,k-1}, R_u, E_x)$ with $E_x = CQC^T + R_x$ is the covariance matrix of $Cw_{k-1} + v_{x,k}$. The estimate covariance of this solution is

$$U_{k-1} = (\mathcal{O}^T R^{-1} \mathcal{O})^{-1} = \begin{bmatrix} P_{x,k-1} & P_{xu,k-1} \\ P_{ux,k-1} & P_{u,k-1} \end{bmatrix}. \quad (14)$$

Therefore, the input estimate $\hat{u}_{k-1}$ is achieved. Since (12) includes the state estimates, measurements from two adjacent time steps, it is less vulnerable to false data injection attacks (FDIA). Other estimation techniques can also be applied to solve (12) to improve outlier robustness such as the least absolute value (LAV), generalized maximum likelihood (GML).

*Bad Data Detection:*

The residual vector and its covariance matrix of the input estimation is given as

$$r_{k-1} = \begin{bmatrix} \hat{x}_{k-1} \\ z_{u,k-1} \\ z_{x,k} \end{bmatrix} - \mathcal{O} \begin{bmatrix} \hat{x}_{k-1} \\ \hat{u}_{k-1} \end{bmatrix}. \quad (15)$$

$$S_{k-1} = \mathcal{O}^T U_{k-1} \mathcal{O} + R. \quad (16)$$

The bad data detection can be performed by checking the Mahalanobis distance $d_M(r_{k-1})$ of the residual vector with a threshold $\zeta$.

$$d_M(r_{k-1}) = \sqrt{r_{k-1}^T S_{k-1}^{-1} r_{k-1}} \geq \zeta \quad (17)$$

The alert is raised when (17) is true. With small $\zeta$, this testing is more sensitive to bad data and attacks, however it also generates more false alarms. Choosing an appropriate $\zeta$ does not discuss in this paper due to limited pages.

*State Estimation*:

The predicted state vector and its covariance can be achieved as

$$\hat{x}_{k|k-1} = A_d \hat{x}_{k-1} + B_d \hat{u}_{k-1}, \quad (18)$$

$$P_{x,k|k-1} = [A_d \quad B_d] U_{k-1} \begin{bmatrix} A_d \\ B_d \end{bmatrix} + Q. \quad (19)$$

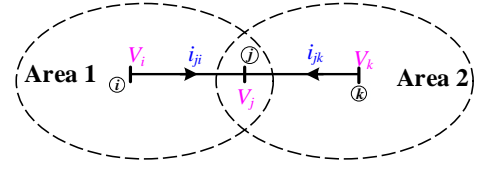

Figure 2. Subsystems partitioning for distributed DSE

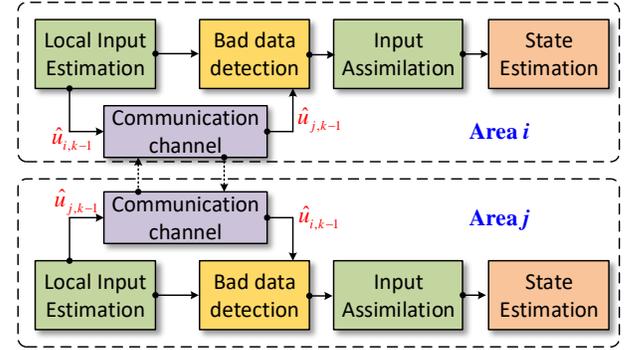

Figure 3. Proposed distributed DSIE and bad data detection

The state estimate and its covariance can be obtained as

$$\hat{x}_k = \hat{x}_{k|k-1} + K_k(z_{x,k} - C\hat{x}_{k|k-1}), \quad (20)$$

$$P_{x,k} = (I - K_k C) P_{k|k-1}. \quad (21)$$

where $K_k = P_{x,k|k-1} C^T (C P_{x,k|k-1} C^T + R_x)^{-1}$ is the updated gain.

III. DISTRIBUTED SCHEME

Any distributed DSE framework includes three stages. First, the power system is partitioned into areas with overlapped regions. Secondly, field distributed controllers perform local DSE using its internal data and measurements. Thirdly, neighbor estimators communicate to exchange and fuse their internal estimates for consistency.

There are some methods introduced for multi-area partitioning of power system [19] provided that each area is observable. In this paper, the partitioning scheme is with the coupled nodal voltages as shown in Fig. 2, where the common buses $j$ are the overlapped regions and its nodal voltages are the shared inputs of adjacent subsystems.

The state-space models of subsystems are built as similar to the centralized model but only based on the local networks and measurements, as follows,

$$\begin{cases} x_{i,k} = A_{id} x_{i,k-1} + B_{id} u_{i,k-1} + w_{i,k-1}, \\ z_{xi,k} = C_i x_k + v_{xi,k}, \quad z_{ui,k} = D_i u_k + v_{ui,k}, \end{cases} \quad (22)$$

where $i$ indicates the subsystems index.

The proposed distributed DSIE and bad data detection are illustrated in Fig. 3. The local input estimation and bad data detection can be performed by adapting (11)-(17) to the local state-space model (19). Then, information of shared inputs is exchanged between neigboring areas through communication channels. The bad data detection of coming information is performed by testing the absolute difference $r_{ij}$ between local and neighbor estimates.

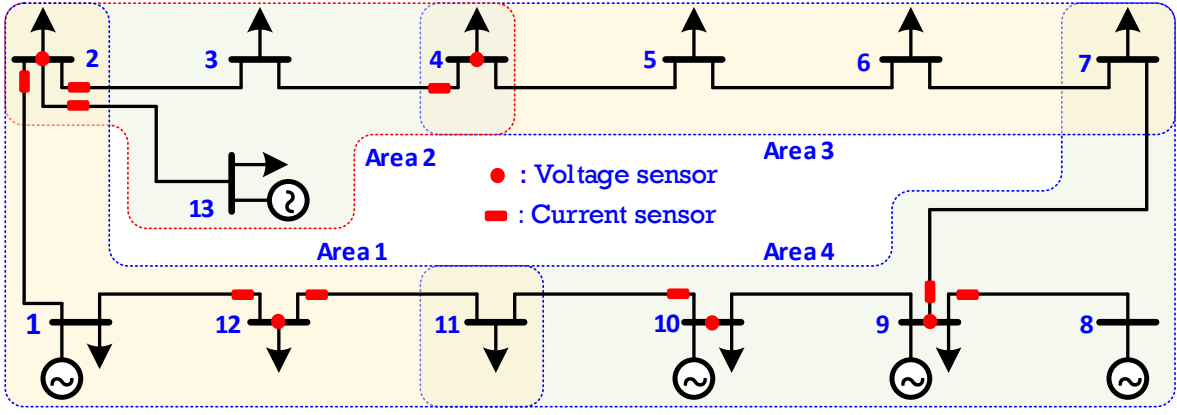

Figure 4. Potsdam (New York) microgrid and multi-area configuration

$$r_{ij,k-1} = |\hat{u}_{i,k-1} - T_{ij}\hat{u}_{j,k-1}|. \quad (23)$$

where $\hat{u}_{j,k-1}$ is the shared input vector, $T_{ij}$ is the transformation matrix so that $u_{j,k-1} = T_{ij}u_{i,k-1}$.

Combining local and neighbor estimates of inputs, the set of equations (24) can be achieved as

$$\begin{bmatrix}\hat{x}_{i,k-1}\\ \hat{u}_{i,k-1}\\ \hat{u}_{j,k-1}\end{bmatrix} = \begin{bmatrix}I & 0\\ 0 & I\\ 0 & T_{ij}\end{bmatrix}\begin{bmatrix}x_{i,k-1}\\ u_{i,k-1}\end{bmatrix} \quad (24)$$

The covariance matrix of $[\hat{x}_{i,k-1}\ \hat{u}_{i,k-1}\ \hat{u}_{j,k-1}]^T$ is $\begin{bmatrix}U_{i,k-1} & 0\\ 0 & P_{uj,k-1}\end{bmatrix}$. The shared input estimates from neighbor areas are considered as new measurements and then are updated to the local estimates by solving (24).

The local state estimation is performed using (18)-(21) but the input and previous state estimates are replaced by those achieved from solving (24).

## IV. SIMULATION RESULTS & DISCUSSIONS

The Potsdam (New York) 13-bus microgrid network is simulated in MATLAB-Simulink and the proposed DSIE framework implemented. Fig. 4 shows the multi-area configuration of this power networks with buses as the shared inputs. The micro-phasor measurement units (µPMU) of voltage and current measurements are denoted as the red dot and rectangles, respectively. The process and measurements are assumed as white and having zero mean, Gaussian distribution with variances of 0.075% and 0.1% respectively. The system operates with load changes every 0.1 s to generate fluctuations in voltages and currents.

The simulation results compare the WLS method, the proposed dynamic SIE method, and the tracking estimation method (TSE) [4]. Fig. 5 shows the local estimates of $v_{3d}$ in areas 2. Fig. 6 shows the local and fused estimates of $v_{11d}$ in areas 1 and 4. Fig. 7 compares the mean squared errors of three methods. It is evident that the proposed dynamic SIE method

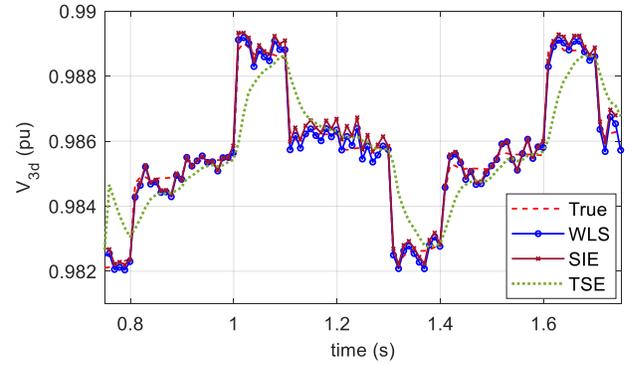

Figure 5. The comparison of estimates of $v_{3d}$ in areas 2

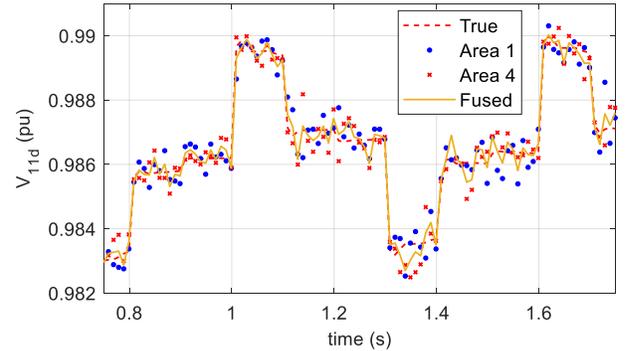

Figure 6. The comparison of estimates of $v_{11d}$ in areas 1 and area 4.

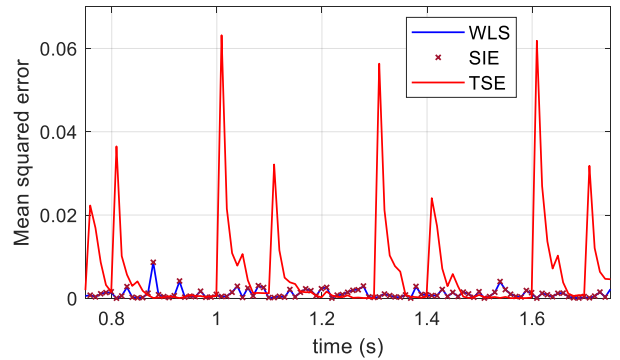

Figure 7. The comparison of mean squared errors.

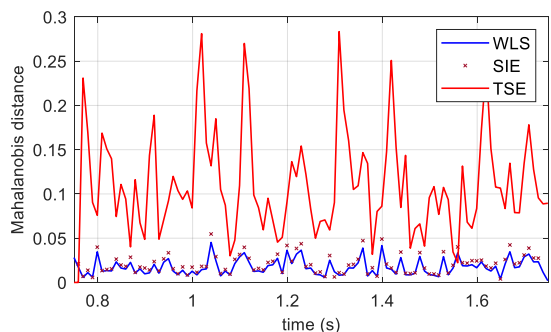

Figure 8. Mahalanobis distances under normal operation.

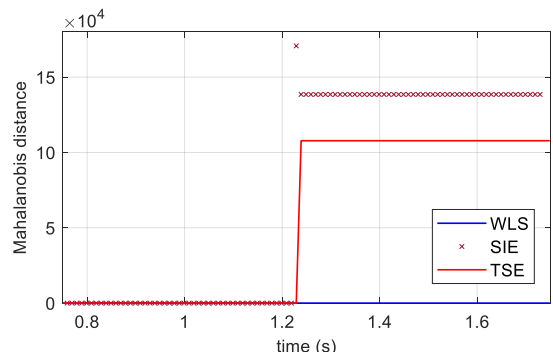

Figure 9. Mahalanobis distances under attacks.

have the performance as the WLS method and these two are much better than the TSE method.

Fig. 8 shows the Mahalanobis distances of residual vectors under normal operation. As can be seen in this figure, the proposed dynamic SIE and WLS method yield quite small distances, while the TSE generates higher distances, especially at load change moments. Fig. 9 shows the Mahalanobis distances of residual vectors under an attack vector is inserted into measurements. Apparently, the WLS method fails to detect the attack since its Mahalanobis distance did not change. The proposed dynamic SIE and TSE methods are able to identify this attack owing to the huge rise of their Mahalanobis distances.

## V. CONCLUSION

In this paper, a distributed scheme for dynamic state and input estimation of microgrids and active distribution networks is proposed. By using the differential equations of power networks and considering sources and loads as inputs, the dynamic state-space model is built and utilized for estimation formulation. A novel Kalman-based filtering method is derived to estimate both states and inputs simultaneously. By comparing with traditional WLS and TSE under normal operation and an attack case, the simulation results have been demonstrated the feasibility and advantages of proposed methods.

The future work will be on incorporating the power measurements and modeling of inverter-based DGU and ESS.

ACKNOWLEDGEMENT

This material is based upon research supported by, or in part by, the U.S. Office of Naval Research under award number N00014-16-1-2956.

REFERENCES

[1] A. Monticelli, *State Estimation in Electric Power Systems*. Boston, MA: Springer US, 1999.

[2] F. C. Schweppe and J. Wildes, "Power System Static-State Estimation, Part I: Exact Model," *IEEE Trans. Power Appar. Syst.*, vol. PAS-89, no. 1, pp. 120–125, 1970.

[3] D. Simon, *Optimal State Estimation*. Hoboken, NJ, USA: John Wiley & Sons, Inc., 2006.

[4] J. Zhao *et al.*, "Power System Dynamic State Estimation: Motivations, Definitions, Methodologies, and Future Work," *IEEE Trans. Power Syst.*, vol. 34, no. 4, pp. 3188–3198, 2019.

[5] D. Ding, Q. L. Han, Z. Wang, and X. Ge, "A Survey on Model-Based Distributed Control and Filtering for Industrial Cyber-Physical Systems," *IEEE Trans. Ind. Informatics*, vol. 15, no. 5, pp. 2483–2499, 2019.

[6] C. Y. Chong, K. C. Chang, and S. Mori, "A Review of Forty Years of Distributed Estimation," *2018 21st Int. Conf. Inf. Fusion, FUSION 2018*, no. September, pp. 70–77, 2018.

[7] T. Van Cutsem, J. Horward, and M. Ribbens-Pavella, "A Two-Level Static State Estimator for Electric Power Systems," *IEEE Trans. Power Appar. Syst.*, vol. PAS-100, no. 8, pp. 3722–3732, Aug. 1981.

[8] A. K. Singh and B. C. Pal, "Decentralized dynamic state estimation in power systems using unscented transformation," *IEEE Trans. Power Syst.*, vol. 29, no. 2, pp. 794–804, 2014.

[9] A. Benigni, G. D'Antona, U. Ghisla, A. Monti, and F. Ponci, "A decentralized observer for ship power system applications: Implementation and experimental validation," *IEEE Trans. Instrum. Meas.*, vol. 59, no. 2, pp. 440–449, 2010.

[10] C. Wang, Z. Qin, Y. Hou, and J. Yan, "Multi-area dynamic state estimation with PMU measurements by an equality constrained extended kalman filter," *IEEE Trans. Smart Grid*, vol. 9, no. 2, pp. 900–910, 2018.

[11] M. M. Rana, L. Li, S. W. Su, and W. Xiang, "Consensus-based smart grid state estimation algorithm," *IEEE Trans. Ind. Informatics*, vol. 14, no. 8, pp. 3368–3375, 2018.

[12] M. Rostami and S. Lotfifard, "Distributed dynamic state estimation of power systems," *IEEE Trans. Ind. Informatics*, vol. 14, no. 8, pp. 3395–3404, 2018.

[13] S. J. Geetha, S. Chakrabarti, K. Rajawat, and V. Terzija, "An Asynchronous Decentralized Forecasting-Aided State Estimator for Power Systems," *IEEE Trans. Power Syst.*, vol. 34, no. 4, pp. 3059–3068, 2019.

[14] S. Riverso, F. Sarzo, and G. Ferrari-Trecate, "Plug-and-Play Voltage and Frequency Control of Islanded Microgrids with Meshed Topology," *IEEE Trans. Smart Grid*, vol. 6, no. 3, pp. 1176–1184, 2015.

[15] L. H. Bang Nguyen, T. V. Vu, and T. A. Ngo, "Decentralized Dynamic State Estimation in Microgrids," in *2019 IEEE Electric Ship Technologies Symposium (ESTS)*, 2019, pp. 257–262.

[16] D. A. Haughton and G. T. Heydt, "A linear state estimation formulation for smart distribution systems," *IEEE Trans. Power Syst.*, vol. 28, no. 2, pp. 1187–1195, 2013.

[17] I. Markovsky and B. De Moor, "Linear dynamic filtering with noisy input and output," *Automatica*, vol. 41, no. 1, pp. 167–171, 2005.

[18] S. Gillijns and B. De Moor, "Unbiased minimum-variance input and state estimation for linear discrete-time systems with direct feedthrough," *Automatica*, vol. 43, no. 5, pp. 934–937, 2007.

[19] A. Gómez-Expósito, A. De La Villa Jaén, C. Gómez-Quiles, P. Rousseaux, and T. Van Cutsem, "A taxonomy of multi-area state estimation methods," *Electr. Power Syst. Res.*, vol. 81, no. 4, pp. 1060–1069, 2011.